\documentclass[twocolumn,english,aps,prb,floats]{revtex4-2}
\usepackage[T1]{fontenc}
\usepackage[latin9]{inputenc}
\usepackage{bm}
\usepackage{amsmath}
\usepackage{amssymb}
\usepackage{graphicx}
\usepackage{esint}

\makeatletter
\usepackage{babel}

\makeatother

\usepackage{babel}
\begin{document}
\title{Static and Microwave Properties of Amorphous Magnets Near Saturation}
\author{Eugene M. Chudnovsky and Dmitry A. Garanin}
\affiliation{Physics Department, Herbert H. Lehman College and Graduate School,
The City University of New York, 250 Bedford Park Boulevard West,
Bronx, New York 10468-1589, USA }
\date{\today}
\begin{abstract}
Static and dynamic properties of magnetically soft amorphous ferromagnets
have been studied analytically and numerically within random-field
and random-anisotropy models. External field and coherent anisotropy
that are weak compared to their random counterparts are sufficient
to bring the magnet close to saturation. The scaling of spin-spin
correlations in this regime is computed, and its dependence on parameters
is confirmed by Monte Carlo simulation. We show that near the ferromagnetic
resonance, the spin excitations are damped and spatially localized
due to randomness even close to saturation. On increasing the strength
of randomness, the localization length goes down in accordance with
theoretical expectations, while the damping of spin excitations goes
up. 
\end{abstract}
\maketitle

\section{Introduction}

\label{Sec_Intro}

\begin{figure}[h]
\centering{}\includegraphics[width=8cm]{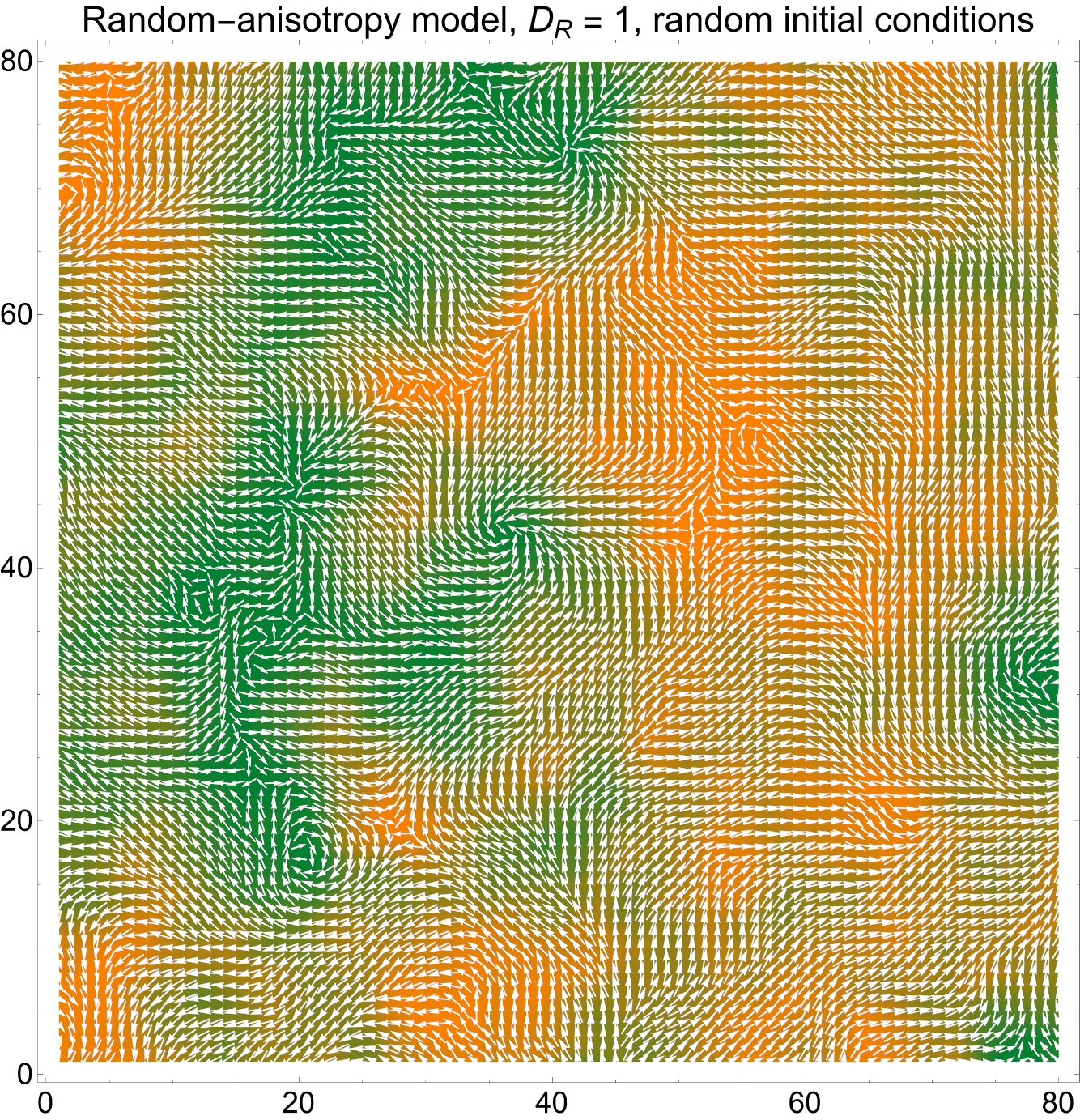}\caption{Equilibrium spin structure in a 2D RA model for $D_{R}/J=1$ and $T=0$
obtained numerically by the energy minimization starting from a random
initial condition. In-plane spin components are shown by white arrows.
The out-of-plane component is shown by orange/green corresponding
to positive/negative. Axes labels are distances in lattice units.}
\label{Fig_RA_structure} 
\end{figure}

Static properties of amorphous magnets have been studied theoretically
within the random field (RF) and random anisotropy (RA) models \citep{key-1,EC-1983,AP-PRB1983,CSS-1986,RA-book,Fisch1998,Fisch2000,CT-book,Fisch2007,garchupro2013prb,PGC-PRL,garchu15epjb,PCG-2015,CG-PRL,GC-JPhys2022}
since the early 1980s. Both models assume a ferromagnetic exchange
between neighboring spins and a random field or random magnetic anisotropy
acting on each spin. The behavior of the magnetization in an amorphous
magnet is illustrated by Fig.\ \ref{Fig_RA_structure}. It was first
understood by Imry and Ma \citep{IM} for arbitrary systems with a
continuous order parameter and quenched randomness. In application
to disordered magnets, it comes from the competition between random
local forces rotating the spins and the ferromagnetic exchange that
favors their uniform alignment. This competition results in a finite
ferromagnetic correlation length that increases on decreasing the
strength of the local disordering forces. 

Theoretical work on the dynamical properties of the RF and RA models
has been scarce, mainly focused \citep{Saslow2018} on the phenomenology
of the ferromagnetic resonance (FMR) investigated experimentally \citep{Monod,Prejean,Alloul1980,Schultz,Gullikson,Sheftel2018}.
In accordance with general theoretical arguments \citep{Fert,Levy,Henley1982,HS-1977,Saslow1982,Bruinsma1986,Serota1988,Ma-PRB1986,Zhang-PRB1993,Alvarez-PRL2013,Yu-AnnPhys2013,Nowak2015},
localization of spin modes was observed in experiments on disordered
magnets \citep{Amaral-1993,Suran1-1997,Suran2-1997,Suran-1998,McMichael-PRL2003,Loubens-PRL2007,Du-PRB2014}.
More recently, localized spin-wave excitations generated by microwaves
in the RA system have been studied on large spin lattices numerically
\citep{GC-PRB2021,GC-localization}. It was demonstrated \citep{GC-PRB2021,GC-PRB2022,GC-localization,CG-integral,EC-DG-scaling,EC-DG-dyn-scaling}
that random-anisotropy magnets can be promising materials for broadband
microwave absorption. This material property is pursued due to its
numerous technological applications such as, e.g., microwave shielding,
thermal cancer treatment, and stealth technology \citep{carbon2020}.
Nanoscale inhomogeneity of the magnetization in amorphous random-anisotropy
magnets \citep{PSS2016,Martin2020}, see Fig.\ \ref{Fig_RA_structure},
resembles nanocomposites \citep{nanocomposites,AFM2023} with embedded
magnetic particles but has an advantage of a greater magnetic volume
available for microwave absorption.

In this paper, we study within microscopic models the static and dynamic
properties of RF and RA magnets close to saturation caused by either
the external magnetic field or coherent (global) magnetic anisotropy.
The importance of this problem becomes clear from the magnetic softness
of such magnets. As we shall see, a magnetic field that is weak compared
to the RF, or the coherent anisotropy (CA) that is weak compared to
the RA, is sufficient to bring the magnet close to saturation. Some
CA would inevitably be present in an amorphous magnet due to the manufacturing
process. Thus the static and dynamic properties of the CA+RA model,
including microwave absorption, require special consideration, as
compared to the pure RA model \citep{GC-PRB2021,GC-PRB2022,GC-localization,CG-integral,EC-DG-scaling,EC-DG-dyn-scaling}.

The characteristic scale of the spatial fluctuations of the magnetization
is the ferromagnetic correlation length. In RF and RA models it corresponds
to the average size of Imry-Ma domains due to randomness illustrated
by Fig.\ \ref{Fig_RA_structure}. When it is small compared to the
size of conventional magnetic domains originating from magnetic dipolar
forces, or when the size of the system is small compared to the typical
size of such domains, the magnetic dipole-dipole interaction between
the spins can be neglected. Here we consider such a situation.

Both problems, RF and RA, are highly nonlinear. However, the RF problem
is easier to tackle analytically. While it is more theoretical than
practical, it serves as a precursor for the RA problem. For a weak
RF it allows one to obtain rigorous scaling of physical quantities
with parameters. Then, using the IM argument, and the language of
the anisotropy field, that scaling can be translated into the dependence
of the physical quantities on the CA+RA model on the parameters. The
latter can be tested in a numerical experiment, which will be done
here.

The paper is organized as follows. Static spin-spin correlations are
studied for the three-component spins in three (bulk solid) and two
(thin film) dimensions in Section \ref{Sec_Random=000020magnet}.
This is done analytically within the RF model, first without an external
field, Section \ref{Sec_spin_correlations_RF_no_H}, and then analytically
in the presence of the external field, Section \ref{Sec_spin=000020correlations_RF_with_H}.
The obtained scaling is extended to the dependence of static spin-spin
correlations on parameters in the CA+RA model, which is confirmed
numerically in Section \ref{Sec_spin_correlations_RA+CA}. Dynamical
properties close to saturation are studied in Section \ref{Sec_dynamics}.
First, the analytical theory of spin excitations in the RF model with
the external field is presented in Section \ref{Sec_excitations-analytical}.
Then excitations in the CA+RA model are studied numerically in Section
\ref{Sec_excitations_numerical} and their localization length is
computed. Section \ref{Sec_precession_decay} illustrates the dynamics
of the system in the presence of RA that causes damping of the uniform
precession. Our conclusions are outlined in Section \ref{Sec_conclusions}.

\section{Ferromagnet with weak static randomness close to saturation}

\label{Sec_Random=000020magnet}

We shall consider two models, the random-field (RF) model with the
Hamiltonian 
\begin{equation}
{\cal H}=-\frac{1}{2}\sum_{ij}J_{ij}{\bf s}_{i}\cdot{\bf s}_{j}-\sum_{i}{\bf h}_{i}\cdot{\bf s}_{i}-{\bf H}\cdot\sum_{i}{\bf s}_{i},\label{eq:ham-h-discrete}
\end{equation}
and the random anisotropy (RA) model with the Hamiltonian 
\begin{equation}
{\cal H}=-\frac{J}{2}\sum_{ij}{\bf s}_{i}\cdot{\bf s}_{j}-\frac{D_{R}}{2}\sum_{i}({\bf n}_{i}\cdot{\bf s}_{i})^{2}-\frac{D_{C}}{2}\sum_{i}({\bf n}_{c}\cdot{\bf s}_{i})^{2},\label{eq:ham-RA-discrete}
\end{equation}
Here ${\bf s}_{i}$ is a $n$-component constant-length ($\left|{\bf s}_{i}\right|=s$)
spin at the site $i$ of a cubic lattice. The summation is over the
nearest neighbors. In Eq.\ (\ref{eq:ham-h-discrete}) ${\bf h}_{i}$
is a static random field at a site $i$, while in Eq.\ (\ref{eq:ham-RA-discrete})
${\bf n}_{i}$ represents the direction of random magnetic anisotropy
at a site $i$, with $D_{R}$ being its strength. In the first equation
${\bf H}$ is the external magnetic field, while in the second equation
$D_{C}$ is the strength of the coherent magnetic anisotropy having
the same direction ${\bf n}_{c}$ at each lattice site.

While these two models are mathematically different, they are conceptually
similar. Both RF and RA satisfying the condition $h,D_{R}\ll J$ slightly
perturb directions of the spins away from the parallel alignment that
is preferred by the strong ferromagnetic exchange, while both the
external magnetic field and the coherent anisotropy (CA) provide the
preferred direction of the global magnetization. The RF model is easier
to treat analytically, which we will do first. The scaling of physical
parameters on $h/J$ and $H/J$ will be established. Then the RA model
will be studied numerically and its equivalence to the RF model in
terms of the scaling on $D_{R}/J$ and $D_{C}/J$ will be confirmed.

\subsection{Spin correlations in the RF model at ${\bf H}=0$: Analytical theory}

\label{Sec_spin_correlations_RF_no_H}

The continuous counterpart of the RF model, Eq.\ (\ref{eq:ham-h-discrete}),
is described by the Hamiltonian 
\begin{equation}
{\cal H}=\int d^{d}r\left[\frac{\alpha}{2}(\nabla{\bf S})^{2}-{\bf h}\cdot{\bf S}-{\bf H}\cdot{\bf S}\right].\label{H-continuous}
\end{equation}
Here ${\bf S}$ is a spin field of constant length $S=s/a^{d}$ in
$d$ spatial dimensions and $\alpha=Ja^{d+2}$ is the exchange stiffness,
with $a$ being the lattice spacing. We use uncorrelated RF satisfying
\begin{equation}
\left\langle h_{i\alpha}h_{j\beta}\right\rangle =\frac{h^{2}}{n}\delta_{\alpha\beta}\delta_{ij},\label{eq:h-corr-ij}
\end{equation}
for the lattice model and 
\begin{equation}
\langle h_{\alpha}({\bf r}')h_{\beta}({\bf r}'')\rangle=\frac{h^{2}}{n}\delta_{\alpha\beta}a^{3}\delta({\bf r}'-{\bf r}'')\label{eq:h-corr_continuous}
\end{equation}
for the continuous model, with Greek indices being the Cartesian components
of the vectors and $\left|\mathbf{h}_{i}\right|=h=\mathrm{const}$.
Within this model, one obtains in 3D in the absence of the external
field \citep{garchu15epjb} 
\begin{equation}
\frac{1}{2S^{2}}\langle[{\bf S}({\bf r}_{1})-{\bf S}({\bf r}_{2})]^{2}\rangle=\frac{|{\bf r}_{1}-{\bf r}_{2}|}{R_{f}}
\end{equation}
for $|{\bf r}_{1}-{\bf r}_{2}|\ll R_{f}$ with 
\begin{equation}
\frac{R_{f}}{a}=\frac{8\pi\alpha_{e}^{2}S^{2}}{h^{2}a^{4}(1-1/n)}=\frac{8\pi}{(1-1/n)}\left(\frac{Js}{h}\right)^{2}.\label{Rf-n}
\end{equation}
Noticing that 
\begin{equation}
\frac{1}{2S^{2}}\langle[{\bf S}({\bf r}_{1})-{\bf S}({\bf r}_{2})]^{2}\rangle=1-\frac{1}{S^{2}}\langle{\bf S}({\bf r}_{1})\cdot{\bf S}({\bf r}_{2})\rangle
\end{equation}
this gives 
\begin{equation}
\langle{\bf s}_{i}\cdot{\bf s}_{j}\rangle=s^{2}\left(1-\frac{|{\bf r}_{i}-{\bf r}_{j}|}{R_{f}}\right).\label{sR}
\end{equation}
Based upon this result, and assuming that spin-spin correlations in
the RF model at $H=0$ decay exponentially (as suggested by the IM
argument and confirmed by the numerical analysis \citep{PGC-PRL})
one can use 
\begin{equation}
\langle{\bf s}_{i}\cdot{\bf s}_{j}\rangle=s^{2}\exp\left(-\frac{|{\bf r}_{i}-{\bf r}_{j}|}{R_{f}}\right)\label{exp-sR}
\end{equation}
as a good approximation for arbitrary $|{\bf r}_{i}-{\bf r}_{j}|$.
Similar consideration in 2D gives $R_{f}/s\propto Js/h$.

\subsection{Spin correlations in the RF model near saturation: Analytical theory}

\label{Sec_spin=000020correlations_RF_with_H}

We shall now turn to the RF model with a nonzero external field ${\bf H}=H\hat{z}$.
To account for the constant length of the spin field, we consider
the Hamiltonian 
\begin{equation}
{\cal H}_{\lambda}=\int d^{d}r\left[\frac{\alpha}{2}(\nabla{\bf S})^{2}-{\bf h}\cdot{\bf S}-{\bf H}\cdot{\bf S}-\lambda({\bf r}){\bf S}^{2}\right]\label{H-continuous_lambda}
\end{equation}
with the Lagrange multiplier $\lambda({\bf r})$. Properties of this
model near saturation have been previously studied for $XY$ spins
in three dimensions\citep{garchupro2013prb}. We shall now extend
this study to the three-component spins in three and two dimensions.

The extremal field ${\bf S}({\bf r})$ corresponding to the Hamiltonian
(\ref{H-continuous_lambda}) satisfy 
\begin{equation}
\alpha_{e}\nabla^{2}{\bf S}+{\bf H}+{\bf h}+2\lambda{\bf S}=0.\label{extremal}
\end{equation}
with $\lambda$ provided by the condition ${\bf S}^{2}=S^{2}$: 
\begin{equation}
\lambda=-\frac{1}{2S^{2}}(\alpha_{e}{\bf S}\cdot\nabla^{2}{\bf S}+{\bf S}\cdot{\bf H}+{\bf S}\cdot{\bf h})
\end{equation}
This gives 
\begin{equation}
\nabla^{2}{\bf S}-\frac{1}{S^{2}}\left({\bf S}\cdot\nabla^{2}{\bf S}\right){\bf S}-\frac{1}{\alpha_{e}S^{2}}({\bf S}\cdot{\bf H}+{\bf S}\cdot{\bf h}){\bf S}=-\frac{1}{\alpha_{e}}({\bf H}+{\bf h}).\label{equilibrium}
\end{equation}
Close to the saturation, one can write 
\begin{equation}
{\bf S}={\bf S}_{0}({\bf r})=S\hat{z}+{s}_{z}\hat{z}+{\bf s}_{\perp}\quad{\rm with}\quad|s_{\perp}|,|s_{\parallel}|\ll S.
\end{equation}
Here ${s}_{z}\hat{z}$ and ${\bf s}_{\perp}$ are oscillating (in
space) components of the spin field due to the random field. Linearization
of the above equation then gives 
\begin{eqnarray}
 &  & \nabla^{2}s_{z}\hat{z}+\nabla^{2}{\bf s}_{\perp}-\nabla^{2}s_{z}\hat{z}-\frac{1}{\alpha_{e}S}Hs_{z}\hat{z}-\frac{1}{\alpha_{e}}h_{z}\hat{z}\nonumber \\
 &  & -\frac{1}{\alpha_{e}S}Hs_{z}\hat{z}-\frac{1}{\alpha_{e}S}H{\bf s}_{\perp}=-\frac{1}{\alpha_{e}}h_{z}\hat{z}-\frac{1}{\alpha_{e}}{\bf h}_{\perp},
\end{eqnarray}
which implies $s_{z}=0$ in the first order on ${\bf h}$ while ${\bf s}_{\perp}$
satisfies 
\begin{equation}
\left(\nabla^{2}-k_{H}^{2}\right){\bf s}_{\perp}=-\frac{1}{Ja^{d+2}}{\bf h}_{\perp}\label{S-perp}
\end{equation}
where 
\begin{equation}
k_{H}^{2}\equiv\frac{H}{JSa^{d+2}}=\frac{H}{Jsa^{2}},\qquad\frac{1}{k_{H}}=R_{H}=a\sqrt{\frac{Js}{H}}.\label{k-H}
\end{equation}
Here we have used the relations $\alpha_{e}=Ja^{d+2},S=s/a^{d}$.

The length $R_{H}$ represents the correlation length describing the
wandering of the transverse component of the magnetization. The close-to-saturation
condition is equivalent to $a\ll R_{H}\ll R_{f}$, where $R_{f}$
is the ferromagnetic correlation length at $H=0$. Since $R_{H}$
scales as $(Js/H)^{1/2}$, while $R_{f}$ scales as $(Js/h)^{2}$
in 3D and as $Js/h$ in 2D, this condition shows that the external
field that is very weak compared to the random field may be sufficient
to saturate the system. This agrees with a well-known experimental
fact that RF (RA) magnets are magnetically very soft. It also shows
that the regime close to saturation covers almost the entire range
of the magnetic field that would be available in the experiment.

Implicit solution of Eq.\ (\ref{S-perp}) is 
\begin{equation}
{\bf s}_{\perp}({\bf r})=-\frac{1}{Ja^{d+2}}\int d^{d}r'G({\bf r}-{\bf r}'){\bf h}_{\perp}({\bf r}').
\end{equation}
where $G(r)$ is the Green function of Eq.\ (\ref{S-perp}), $G(r)=-e^{-k_{H}r}/(4\pi r)$
in 3D and $G(r)=K_{0}(k_{H}r)/(2\pi)$ in 2D, with $K_{0}$ being
the Macdonald function. This gives 
\begin{eqnarray}
 &  & \langle{\bf s}_{\perp}({\bf r}_{1})\cdot{\bf s}_{\perp}({\bf r}_{2})\rangle=\frac{1}{J^{2}a^{2d+4}}\int d^{d}r'\int d^{d}r''\times\nonumber \\
 &  & G({\bf r}_{1}-{\bf r}')G({\bf r}_{2}-{\bf r}')\langle{\bf h}_{\perp}({\bf r}')\cdot{\bf h}_{\perp}({\bf r}'')\rangle.
\end{eqnarray}
Substituting here 
\begin{equation}
\langle{\bf h}_{\perp}({\bf r}')\cdot{\bf h}_{\perp}({\bf r}'')\rangle=\frac{2}{3}h^{2}a^{d}\delta({\bf r}'-{\bf r''}).
\end{equation}
we obtain 
\begin{equation}
\langle{\bf s}_{\perp}({\bf r}_{1})\cdot{\bf s}_{\perp}({\bf r}_{2})\rangle=\frac{2h^{2}}{3J^{2}a^{d+4}}\int d^{d}rG({\bf r}_{1}-{\bf r})G({\bf r}_{2}-{\bf r}).
\end{equation}
In 3D it gives 
\begin{equation}
\langle{\bf s}_{\perp}({\bf r}_{1})\cdot{\bf s}_{\perp}({\bf r}_{2})\rangle=\frac{h^{2}}{12\pi J^{2}a^{7}k_{H}}e^{-k_{H}|{\bf r}_{1}-{\bf r}_{2}|}.
\end{equation}
The magnitude of the fluctuations of ${\bf s}_{\perp}$ is provided
by this expression at ${\bf r}_{1}={\bf r}_{2}$, 
\begin{equation}
\langle{\bf s}_{\perp}^{2}\rangle=\frac{h^{2}}{12\pi J^{2}a^{7}k_{H}}=\left(\frac{2R_{H}}{3R_{f}}\right)S^{2}\ll S^{2}.\label{magnitude}
\end{equation}
On the approach to saturation, in the field range satisfying $a\ll R_{H}\ll R_{f}$,
one has 
\begin{eqnarray}
S_{z} & = & S-\frac{1}{2S}\langle{\bf s}_{\perp}^{2}\rangle\\
1-\frac{S_{z}}{S} & = & \frac{h^{2}}{24\pi J^{2}S^{2}a^{7}k_{H}}=\frac{1}{24\pi}\left(\frac{h}{Js}\right)^{2}\sqrt{\frac{Js}{H}}.
\end{eqnarray}

At very high fields, when $R_{H}$ becomes formally smaller than $a$,
the magnetization law on the approach to saturation changes. In this
case the Laplacian in Eq (\ref{S-perp}) becomes irrelevant and one
has 
\begin{equation}
{\bf s}_{\perp}=\frac{{\bf h}_{\perp}}{Ja^{d+2}k_{H}^{2}}=\frac{{\bf h}_{\perp}}{H}S,\qquad\langle{\bf s}_{\perp}^{2}\rangle=\frac{\langle{\bf h}_{\perp}^{2}\rangle}{H^{2}}S^{2},
\end{equation}
so that 
\begin{equation}
1-\frac{S_{z}}{S}=\frac{1}{3}\left(\frac{h}{H}\right)^{2}.\label{high-H}
\end{equation}

In 2D one has 
\begin{equation}
\langle{\bf s}_{\perp}({\bf r}_{1})\cdot{\bf s}_{\perp}({\bf r}_{2})\rangle=\frac{h^{2}}{6\pi J^{2}a^{6}k_{H}}|{\bf r}_{1}-{\bf r}_{2}|K_{1}({k_{H}|{\bf r}_{1}-{\bf r}_{2}|}).
\end{equation}
Noticing that $K_{1}(x)\rightarrow[\pi/(2x)]^{1}/2\exp(-x)$ at large
$x$ and $K_{1}(x)\rightarrow1/x$ at $x\rightarrow0$, we obtain
the magnitude of the fluctuations of ${\bf s}_{\perp}$ in 2D: 
\begin{equation}
\langle{\bf s}_{\perp}^{2}\rangle=\frac{h^{2}}{6\pi J^{2}a^{6}k_{H}^{2}}\sim\left(\frac{R_{H}}{R_{f}}\right)^{2}S^{2}\ll S^{2}.
\end{equation}
Consequently in 2D one must have in the field range satisfying $a\ll R_{H}\ll R_{f}$
\begin{equation}
1-\frac{S_{z}}{S}=\frac{h^{2}}{12\pi J^{2}S^{2}a^{6}k_{H}^{2}}=\frac{1}{12\pi}\left(\frac{h}{Js}\right)^{2}\left(\frac{Js}{H}\right)
\end{equation}
on the approach to saturation. At very high fields, the approach to
saturation in 2D is given by the same law, Eq.\ (\ref{high-H}),
as in 3D.

\begin{figure}[h]
\begin{centering}
\includegraphics[width=8cm]{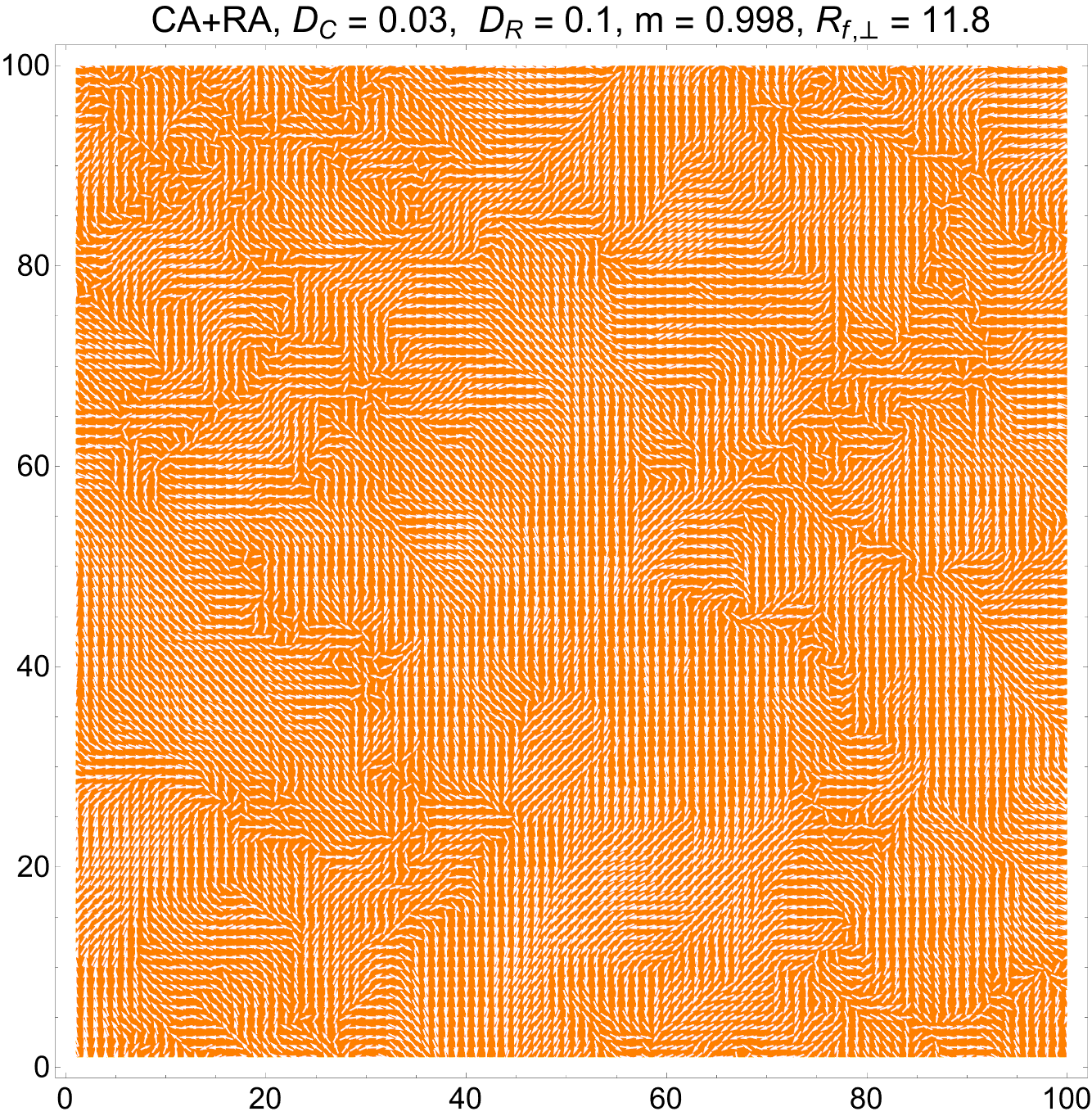}
\par\end{centering}
\centering{}\includegraphics[width=8cm]{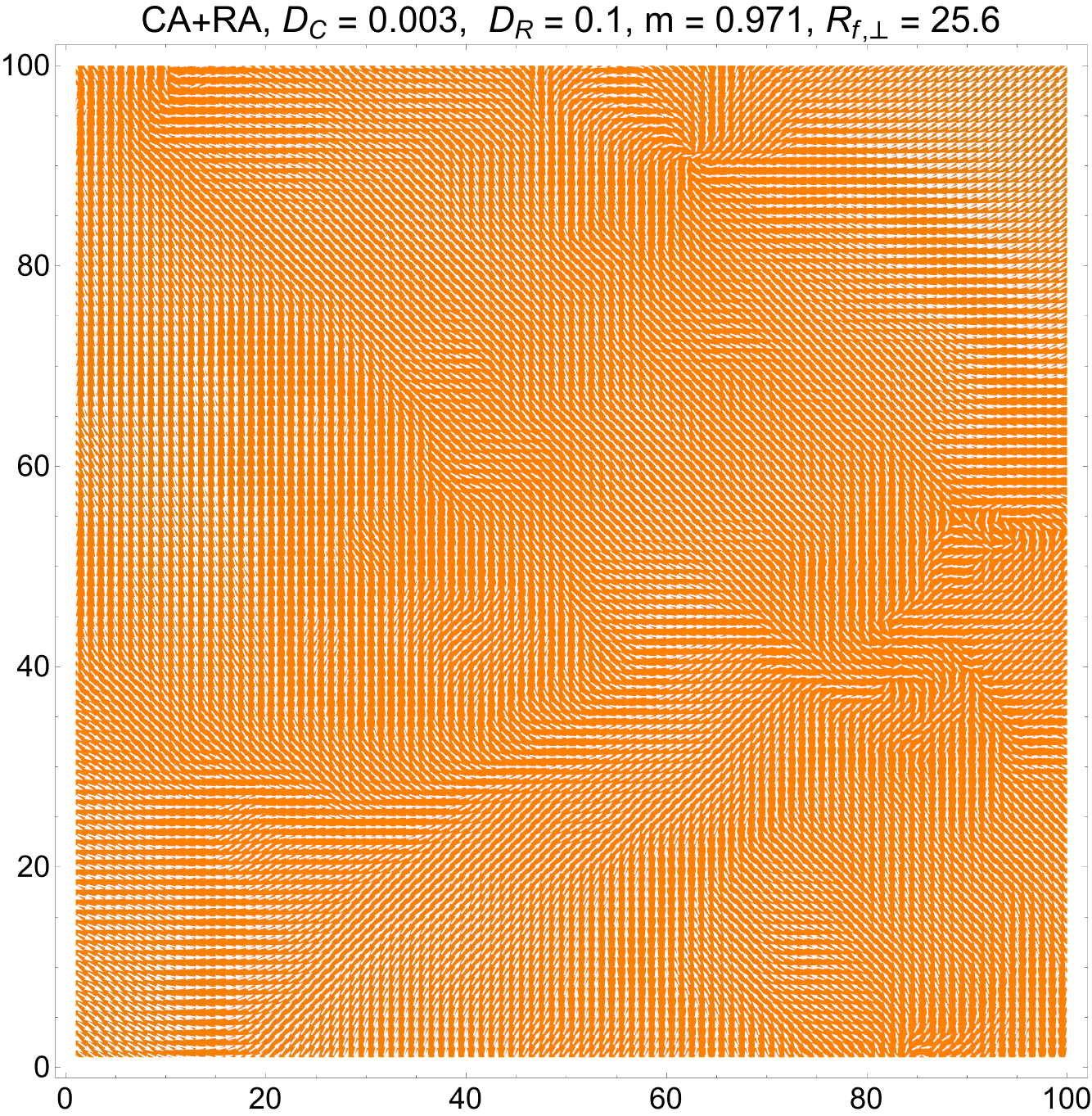}\caption{Transverse spin components (white arrows) in a 2D RA model in the
presence of coherent anisotropy. The transversal correlation length
increases as the coherent anisotropy decreases from $D_{C}/J=0.03$
(upper panel) to $D_{C}/J=0.003$ (lower panel).}
\label{Fig_CA+RA_structure}
\end{figure}

Notice that in all dimensions the crossover to the high-field regime,
$1-S_{z}/S=(1/3)(h/H)^{2}$, occurs at the exchange field $H\sim4\pi Js$.
Such a field would hardly be available in amorphous ferromagnets with
an atomic disorder. However in disordered ferromagnets with large
amorphous structure factor or sintered ferromagnets with $R_{a}\gg a$,
the effective exchange interaction per spin of the grain becomes $J(a/R_{a})^{2}$.
Consequently, the corresponding effective exchange field, $H\sim4\pi Js(a/R_{a})^{2}$,
at which the crossover occurs from the $1/\sqrt{H}$ regime in 3D
or $1/H$ in 2D to the $1/H^{2}$ high-field regime on approach to
saturation would be within experimental reach.

\subsection{Spin correlations in the RA model with coherent anisotropy: Numerical
experiment}

\label{Sec_spin_correlations_RA+CA}

\begin{figure}[h]
\centering{}\includegraphics[width=8cm]{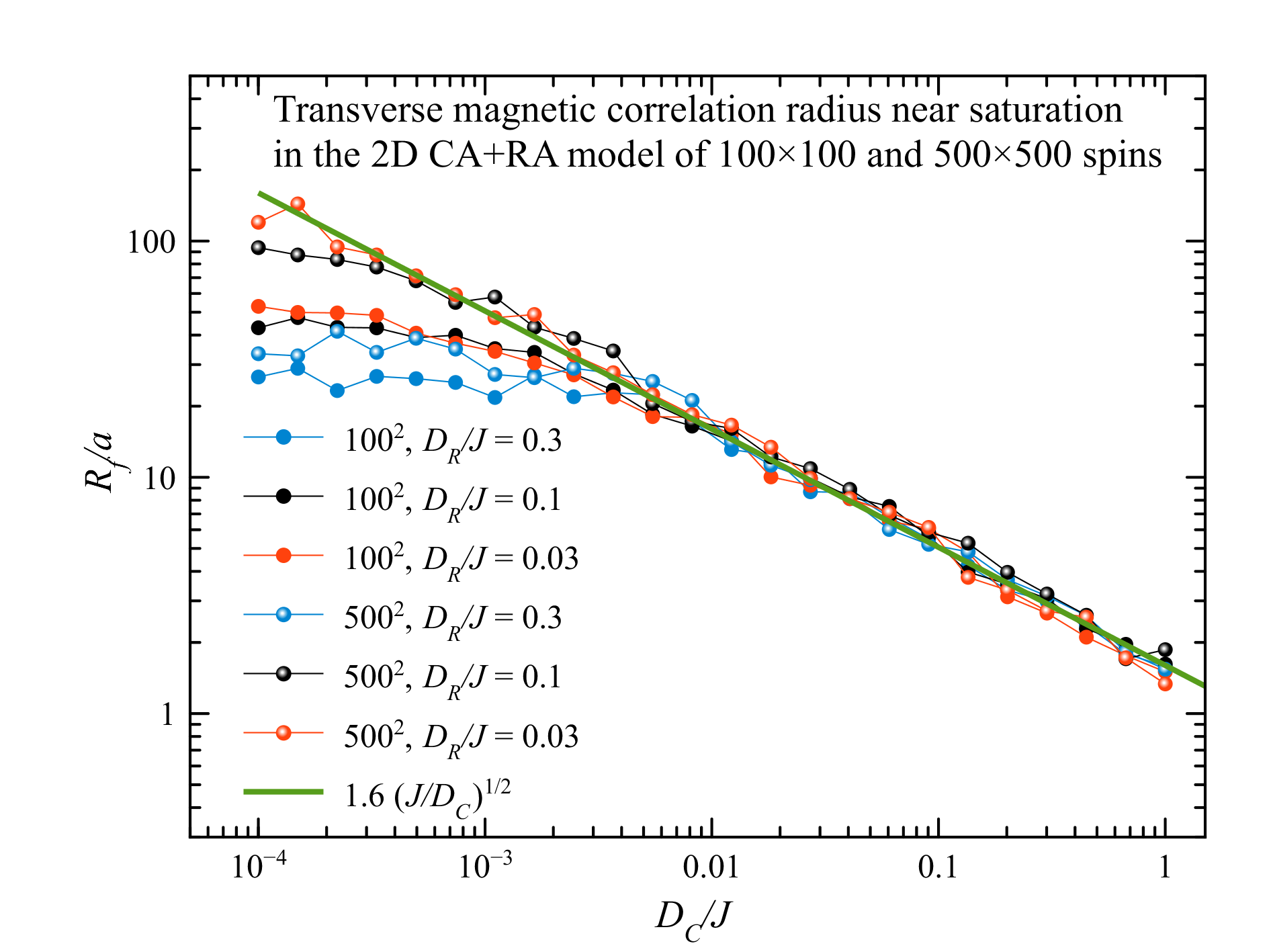}\caption{Dependence of the transversal correlation length $R_{f,\perp}$ on
coherent anisotropy $D_{C}$ in a 2D RA model. It follows theoretical
prediction $R_{f,\perp}\propto(J/D_{C})^{1/2}$ as soon as $R_{f,\perp}$
becomes small compared to the linear size of the system.}
\label{Fig_Rfperp_vs_DC}
\end{figure}

The RA model has a greater nonlinearity than the RF model and is more
difficult to treat by analytical methods. However, as has been already
mentioned, the two models are similar when described in terms of the
Imry-Ma argument. This is reflected in the concept of the anisotropy
field that is commonly used in magnetism. In the RA model, one should
therefore expect that the ratios $h/J$ and $H/J$ that determine
the behavior of physical quantities in the RF model will be replaced
by the ratios $D_{R}/J$ and $D_{C}/J$ in the RA model.

We prepare the states of the CA+RA system at $T=0$ by the energy
minimization starting from the collinear spin state along $z$ axis
and in these states, we extract the values of the transverse ferromagnetic
correlation radius $R_{f,\perp}$ from the values of the transverse
magnetization components, $m_{x}^{2}+m_{y}^{2}$ assuming a Gaussian
shape of the magnetic correlation functions. Our numerical method
has been described in detail in Refs.\ \onlinecite{GC-PRB2021,GC-PRB2022,GC-localization,EC-DG-scaling}.
It reproduces the analytical results of the RF model. To avoid redundancy,
here we will only present the numerical results for the CA+RA model.
In Fig.\ \ref{Fig_CA+RA_structure} we show the transverse components
of the magnetization in the CA+RA model close to saturation. Weak
CA has been chosen, $D_{C}/J=0.03$ (upper panel) and $D_{C}/J=0.003$
(lower panel) compared to the RA with $D_{R}/J=0.1$. In both cases
the system is close to saturation, 99.8\% for $D_{C}/J=0.03$ and
97.1\% for $D_{C}/J=0.003$. This is in accordance with the expectation
based upon the results of the previous section that $D_{C}\ll D_{R}$
is sufficient to almost saturate the system. Slightly lower magnetization
for weaker CA is expected from the fact that the ordering due to the
CA competes with the disordering due to the RA.

Fig.\ \ref{Fig_Rfperp_vs_DC} shows the dependence of the transversal
correlation length, $R_{f,\perp}$ (that we called $R_{H}$ in the
RF model), on $D_{C}$ for different values of the RA: $D_{R}/J=0.03,0.1,0.3$
in $100\times100$ and $500\times500$ systems. When $R_{\perp}$
becomes small compared to the size of the system, its expected $(J/D_{C})^{1/2}$
scaling, which is independent of $D_{R}$ and system size, is confirmed
numerically with excellent accuracy. This result provides additional
confidence in the conceptual equivalence of the RF and RA models.

\section{Dynamics}

\label{Sec_dynamics}

We shall now consider the dynamics of spin excitations close to saturation.
As before, we will do it analytically for the RF model first, and
then numerically for the CA+RA model. The equivalence of the two models
in terms of the dependence of the physical quantities on parameters
will be demonstrated again. Our most interesting nontrivial observation
is the localization of spin excitations close to saturation where
one would have expected a more conventional uniform ferromagnetic
resonance. We shall see that randomness in the directions of small
transversal components of the magnetization is still sufficient for
localizing excitations.

\subsection{Spin excitations in the RF magnet near saturation: Analytical theory}

\label{Sec_excitations-analytical}

For a random-field ferromagnet, the Landau-Lifshitz equation reads
\begin{equation}
\hbar\frac{\partial{\bf S}}{\partial t}={\bf S}\times{\bf H}_{{\rm eff}},\label{LL}
\end{equation}
where 
\begin{equation}
{\bf H}_{{\rm eff}}=-\frac{\delta\mathcal{H}}{\delta{\bf S}}=J\nabla^{2}{\bf S}+{\bf H}+{\bf h},\label{Heff}
\end{equation}
With an eye on excitations we consider ${\bf S}({\bf r)={\bf S}_{0}({\bf r)+{\bm{\sigma}}({\bf r},t)}}$
where ${\bf S}_{0}=S\hat{z}+{\bf s}_{\perp}({\bf r})$ is the equilibrium
magnetization and ${\bm{\sigma}}({\bf r},t)$ is the time-dependent
perturbation that we choose in the form ${\bm{\sigma}}({\bf r},t)={\bm{\sigma}}({\bf r}){\exp{(-i\omega t)}}$.
The linearized equation of motion is 
\begin{eqnarray}
 &  & -i\hbar\omega{\bm{\sigma}}=Ja^{d+2}S\hat{z}\times{\nabla}^{2}{\bm{\sigma}}+Ja^{d+2}{\bf s}_{\perp}\times{\nabla}^{2}{\bm{\sigma}}\nonumber \\
 &  & +Ja^{d+2}{\bm{\sigma}}\times{\nabla}^{2}{\bm{s}}_{\perp}+{\bm{\sigma}}\times{\bf H}+{\bm{\sigma}}\times{\bf h}.
\end{eqnarray}
Substituting here 
\begin{equation}
Ja^{d+2}\nabla^{2}{\bf s}_{\perp}=Ja^{d+2}k_{H}^{2}{\bf s}_{\perp}-{\bf h}_{\perp}
\end{equation}
from Eq.\ (\ref{equilibrium}), we obtain 
\begin{eqnarray}
 &  & -i\hbar\omega{\bm{\sigma}}=Ja^{d+2}S\hat{z}\times{\nabla}^{2}{\bm{\sigma}}+Ja^{d+2}{\bf s}_{\perp}\times{\nabla}^{2}{\bm{\sigma}}\nonumber \\
 &  & +Ja^{d+2}k_{H}^{2}{\bm{\sigma}}\times{\bm{s}}_{\perp}+H{\bm{\sigma}}\times\hat{z}+h_{z}{\bm{\sigma}}\times\hat{z}.\label{eq-sigma}
\end{eqnarray}
The smallest terms in the right-hand side of the above equation are
$Ja^{d+2}{\bf s}_{\perp}\times{\nabla}^{2}{\bm{\sigma}}+Ja^{d+2}k_{H}^{2}{\bm{\sigma}}\times{\bm{s}}_{\perp}+h_{z}{\bm{\sigma}}\times\hat{z}$.
The first two must contribute negligibly to the time-dependent perturbation
because close to saturation ${\bm{\sigma}}$ is dominated by the $x,y$
components.This allows one to simplify Eq.\ (\ref{eq-sigma}) as
\begin{equation}
-i\hbar\omega{\bm{\sigma}}-Ja^{d+2}S\hat{z}\times{\nabla}^{2}{\bm{\sigma}}-H{\bm{\sigma}}\times\hat{z}=h_{z}{\bm{\sigma}}\times\hat{z}.
\end{equation}
In components one has 
\begin{eqnarray}
-i\hbar\omega{\sigma}_{x}-H\sigma_{y}+Ja^{d+2}S{\nabla}^{2}{\sigma}_{y} & = & h_{z}\sigma_{y}\nonumber \\
-i\hbar\omega{\sigma}_{y}+H\sigma_{x}-Ja^{d+2}S{\nabla}^{2}{\sigma}_{x} & = & -h_{z}\sigma_{x}
\end{eqnarray}
Multiplying the first equation by $i$ and subtracting the second
equation from it, we obtain 
\begin{equation}
Ja^{d+2}S{\nabla}^{2}{\sigma}-(H-\hbar\omega)\sigma=h_{z}\sigma,
\end{equation}
where $\sigma\equiv\sigma_{x}+i\sigma_{y}$.

Switching to three spatial dimensions, it is convenient to write this
last equation in the form 
\begin{equation}
\nabla^{2}\sigma-k_{\omega}^{2}\sigma=\frac{h_{z}}{Jsa^{2}}\sigma,\label{eq-sigma}
\end{equation}
where 
\begin{equation}
k_{\omega}^{2}\equiv\frac{H-\hbar\omega}{Jsa^{2}}.\label{k-omega}
\end{equation}
When the small right-hand side of Eq.\ (\ref{eq-sigma}) is neglected,
one obtains the spin-wave solution with the wave vector $k$ satisfying
$k^{2}=k_{\omega}^{2}$, that is 
\begin{equation}
\hbar\omega=H+Js(ak)^{2},
\end{equation}
as for a conventional ferromagnet. We will see, however, that the
static randomness in the right-hand side of Eq.\ (\ref{eq-sigma})
makes such spin waves localized. Using the 3D Green function of that
equation, $G(r)=-e^{-k_{\omega}r}/(4\pi r)$, one can write its formal
solution as 
\begin{equation}
\sigma({\bf r})=-\frac{1}{4\pi Jsa^{2}}\int d^{3}r'\frac{e^{-k_{\omega}|{\bf r}-{\bf r}'|}}{|{\bf r}-{\bf r}'|}h_{z}({\bf r}')\sigma({\bf r}'),
\end{equation}
so that 
\begin{eqnarray}
 &  & \langle|\sigma({\bf r})|^{2}\rangle=\frac{1}{16\pi^{2}J^{2}s^{2}a^{4}}\int d^{3}r'\int d^{3}r''\times\\
 &  & \frac{e^{-k_{\omega}|{\bf r}-{\bf r}'|}e^{-k_{\omega}|{\bf r}-{\bf r}''|}}{|{\bf r}-{\bf r}'||{\bf r}-{\bf r}''|}\langle h_{z}({\bf r}')h_{z}({\bf r}'')\sigma({\bf r}')\sigma^{*}({\bf r}'')\rangle.\nonumber 
\end{eqnarray}

For long-wavelength oscillations, assuming that $\sigma({\bf r)}$
is correlated on a scale that is much greater than the atomic scale
on which ${h}_{z}({\bf r})$ is correlated, the average under the
integral factorizes: 
\begin{eqnarray}
 &  & \langle h_{z}({\bf r}')h_{z}({\bf r}'')\sigma({\bf r}')\sigma^{*}({\bf r}'')\rangle=\langle h_{z}({\bf r}')h_{z}({\bf r}'')\rangle\langle\sigma({\bf r}')\sigma^{*}({\bf r}'')\rangle\nonumber \\
 &  & =\frac{1}{3}h^{2}a^{3}\delta({\bf r}'-{\bf r}'')\langle\sigma({\bf r}')\sigma^{*}({\bf r}'')\rangle.
\end{eqnarray}
This gives 
\begin{eqnarray}
 &  & \langle|\sigma({\bf r})|^{2}\rangle=\frac{h^{2}}{48\pi^{2}J^{2}s^{2}a}\int d^{3}r'\frac{e^{-2k_{\omega}|{\bf r}-{\bf r}'|}}{|{\bf r}-{\bf r}'|^{2}}\langle|\sigma({\bf r}')|^{2}\rangle\nonumber \\
 &  & =\frac{1}{4\pi R_{f}}\int d^{3}r'\frac{e^{-2k_{\omega}|{\bf r}-{\bf r}'|}}{|{\bf r}-{\bf r}'|^{2}}\langle|\sigma({\bf r}')|^{2}\rangle,\label{eq-integral}
\end{eqnarray}
where we have used Eq.\ (\ref{Rf-n}) for $R_{f}$ in 3D: $R_{f}=12\pi(Js/h)^{2}a$.
Eq.\ (\ref{eq-integral}) is satisfied if 
\begin{equation}
1=\frac{1}{4\pi R_{f}}\int d^{3}r'\frac{e^{-2k_{\omega}|{\bf r}-{\bf r}'|}}{|{\bf r}-{\bf r}'|^{2}}=\frac{2}{k_{\omega}R_{f}},
\end{equation}
which gives $k_{\omega}={2}/{R_{f}}$. With account of Eq.\ (\ref{k-H}),
one obtains 
\begin{equation}
k_{\omega}^{2}=\frac{H-\hbar\omega}{Jsa^{2}}=\frac{4}{R_{f}^{2}},\qquad\hbar\omega=H\left[1-\left(\frac{2R_{H}}{R_{f}}\right)^{2}\right],\label{frequency}
\end{equation}
which is slightly below the FMR frequency due to the disorder and
due to the condition $R_{H}\ll R_{f}$ corresponding to the field
range close to saturation.

An easy check of the self-consistency of the above result comes from
the following consideration based on the conceptual equivalence of
the RF and RA models. As would be expected from the results of Ref.\ \onlinecite{GC-PRB2021},
tested numerically, at $H=0$ the FMR frequency should be determined
by the fluctuation of ${\bf h}$ on the scale $R_{f}$: $\hbar\omega\sim h(a/R_{f})^{3/2}\propto h(h/Js)^{3}$.
We now observe that as $H$ becomes small in Eq.\ (\ref{frequency}),
and $({2R_{H}}/{R_{f}})^{2}$ becomes of order unity, the frequency
of the oscillations, as expected, becomes $\hbar\omega\sim H(R_{H}/R_{f})^{2}\propto h^{4}/(Js)^{3}$.
It is interesting to notice that a small correction $\Delta$ to the
FMR frequency $\omega_{FMR}=H$, which makes the resonance frequency
equal $\omega=H-\Delta$, is independent of $H$. One has $\Delta\sim h^{2}/J$
in 2D, and $\Delta\sim h^{4}/J^{3}$ in 3D.

Switching to two spatial dimensions, the use of the 2D Green function,
$G(r)=K_{0}(k_{\omega}r)/(2\pi)$, modifies the equations as follows
\begin{equation}
\sigma({\bf r})=\frac{1}{2\pi Jsa^{2}}\int d^{2}r'K_{0}[k_{\omega}|{\bf r}-{\bf r}'|]h_{z}({\bf r}')\sigma({\bf r}'),
\end{equation}
\begin{eqnarray}
 &  & \langle|\sigma({\bf r})|^{2}\rangle=\frac{1}{4\pi^{2}J^{2}s^{2}a^{4}}\int d^{2}r'\int d^{2}r''\times\\
 &  & K_{0}[k_{\omega}|{\bf r}-{\bf r}'|]K_{0}[k_{\omega}|{\bf r}-{\bf r}''|]\langle h_{z}({\bf r}')h_{z}({\bf r}'')\sigma({\bf r}')\sigma^{*}({\bf r}'')\rangle.\nonumber 
\end{eqnarray}
With 
\begin{equation}
\langle h_{z}({\bf r}')h_{z}({\bf r}'')\sigma({\bf r}')\sigma^{*}({\bf r}'')\rangle=\langle h_{z}({\bf r}')h_{z}({\bf r}'')\rangle\langle\sigma({\bf r}')\sigma^{*}({\bf r}'')\rangle
\end{equation}
and 
\begin{equation}
\langle h_{z}({\bf r}')h_{z}({\bf r}'')\rangle=\frac{1}{3}\langle h^{2}\rangle a^{2}\delta({\bf r}'-{\bf r''})
\end{equation}
this gives for the three-component spin field and three-component
random field 
\begin{eqnarray}
 &  & \langle|\sigma({\bf r})|^{2}\rangle=\frac{h^{2}}{12\pi^{2}J^{2}s^{2}a^{2}}\int d^{2}r'K_{0}^{2}[k_{\omega}|{\bf r}-{\bf r}'|]\langle|\sigma({\bf r}')|^{2}\rangle\nonumber \\
 &  & \sim\frac{1}{(R_{f}k_{\omega})^{2}}\langle|\sigma({\bf r})|^{2}\rangle,
\end{eqnarray}
where we have used $R_{f}/a\sim Js/h$ in 2D. This again gives $k_{\omega}=1/R_{f}$
and $1-\hbar\omega/H\sim(R_{H}/R_{f})^{2}$. Thus, in all dimensions
$R_{\omega}=1/k_{\omega}$ is independent of the frequency and is
of the order of $R_{f}$ which is the ferromagnetic correlation length
at $H=0$.

For spin excitations of frequency $\omega$ in 3D 
\begin{eqnarray}
 &  & \langle\sigma({\bf r}_{1})\cdot\sigma({\bf r}_{2})\rangle=\frac{1}{16\pi^{2}J^{2}s^{2}a^{4}}\int d^{3}r'\int d^{3}r''\times\nonumber \\
 &  & \frac{e^{-k_{\omega}|{\bf r}_{1}-{\bf r}'|}e^{-k_{\omega}|{\bf r}_{2}-{\bf r}''|}}{|{\bf r}_{1}-{\bf r}'||{\bf r}_{2}-{\bf r}''|}\langle h_{z}({\bf r}')h_{z}({\bf r}'')\sigma({\bf r}')\sigma^{*}({\bf r}'')\rangle\\
 &  & =\frac{\langle|\sigma|^{2}\rangle}{4\pi R_{f}}\int d^{3}r\frac{e^{-k_{\omega}|{\bf r}_{1}-{\bf r}|}}{|{\bf r}_{1}-{\bf r}|^{2}}\frac{e^{-k_{\omega}|{\bf r}_{2}-{\bf r}|}}{|{\bf r}_{2}-{\bf r}|^{2}}=\langle|\sigma|^{2}\rangle e^{-k_{\omega}|{\bf r}_{1}-{\bf r}_{2}|}.\nonumber 
\end{eqnarray}
In 2D one obtains 
\begin{equation}
\langle\sigma({\bf r}_{1})\cdot\sigma({\bf r}_{2})\rangle=\langle|\sigma|^{2}\rangle k_{\omega}|{\bf r}_{1}-{\bf r}_{2}|K_{1}({k_{\omega}|{\bf r}_{1}-{\bf r}_{2}|}).
\end{equation}
This suggests that in the field range corresponding to $a\ll R_{H}\ll R_{F}$,
spin excitations are localized on the scale $R_{\omega}=1/k_{\omega}\sim R_{f}\gg R_{H}$
independently of the frequency and the magnetic field. Such independence
of the localization length on the frequency has been seen in our previous
numerical studies of localization \citep{GC-localization}.

\begin{figure}[h]
\begin{centering}
\includegraphics[width=8cm]{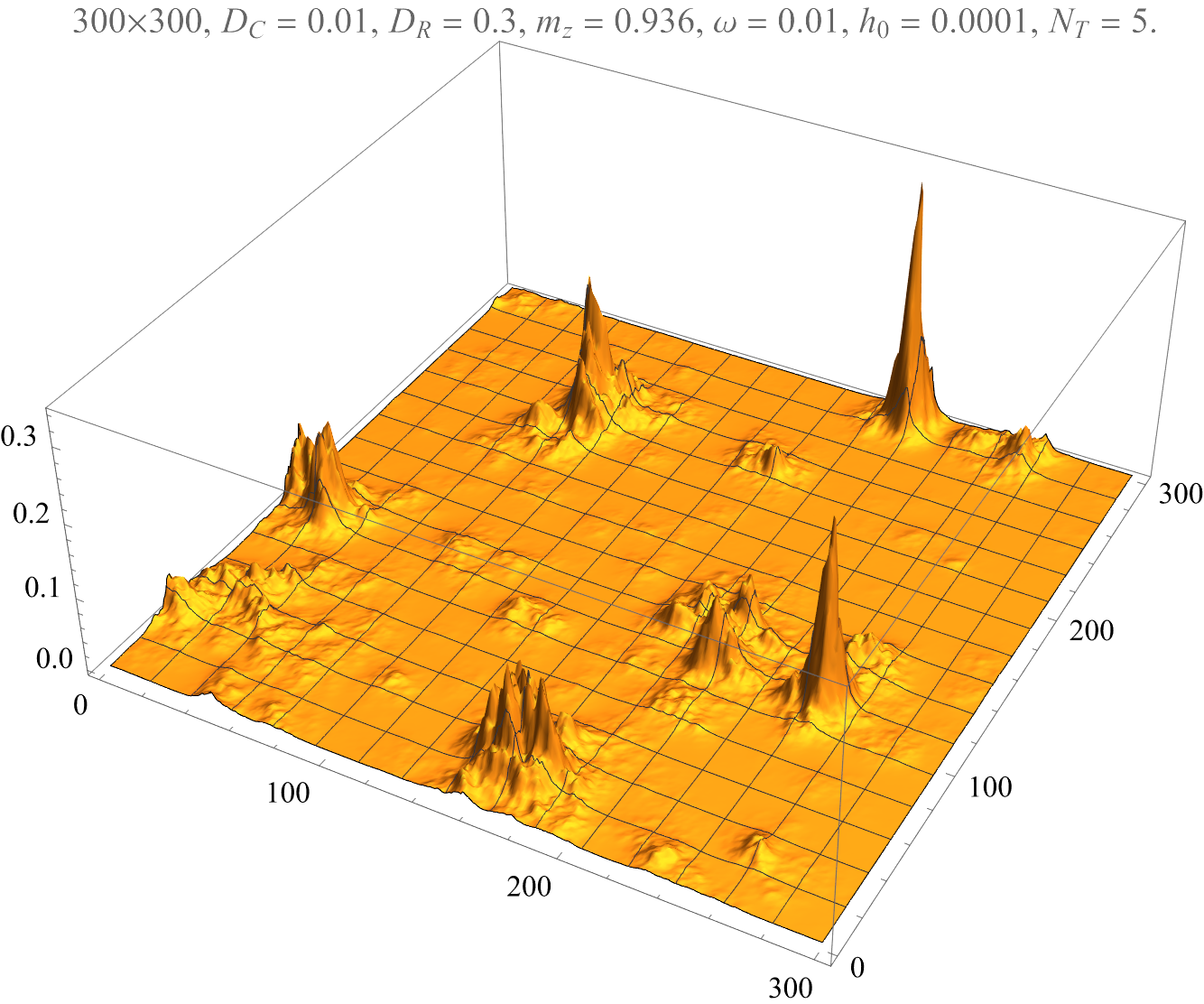}
\par\end{centering}
\centering{}\includegraphics[width=8cm]{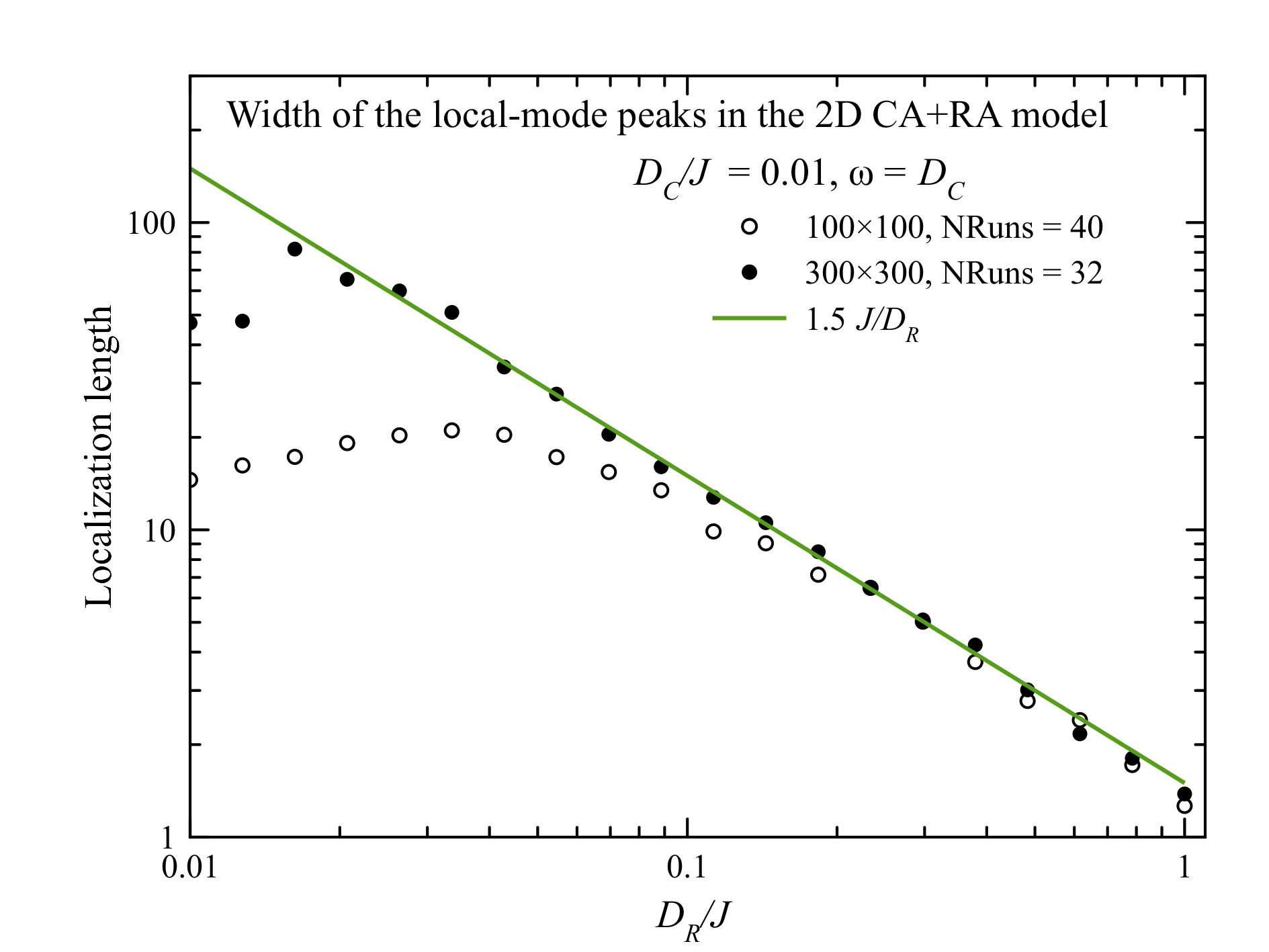}\caption{Upper panel: Localization of spin excitations in a 2D RA model with
coherent anisotropy. Lower panel: Dependence of the localization length
on the RA scales as $J/D_{R}$ in accordance with theoretical prediction
as soon as the peak width becomes small compared to the size of the
system.}
\label{Fig_localization}
\end{figure}

\subsection{Spin excitations in the RA magnet with coherent anisotropy: Numerical
experiment}

\label{Sec_excitations_numerical}

\begin{figure}[h]
\begin{centering}
\includegraphics[width=8cm]{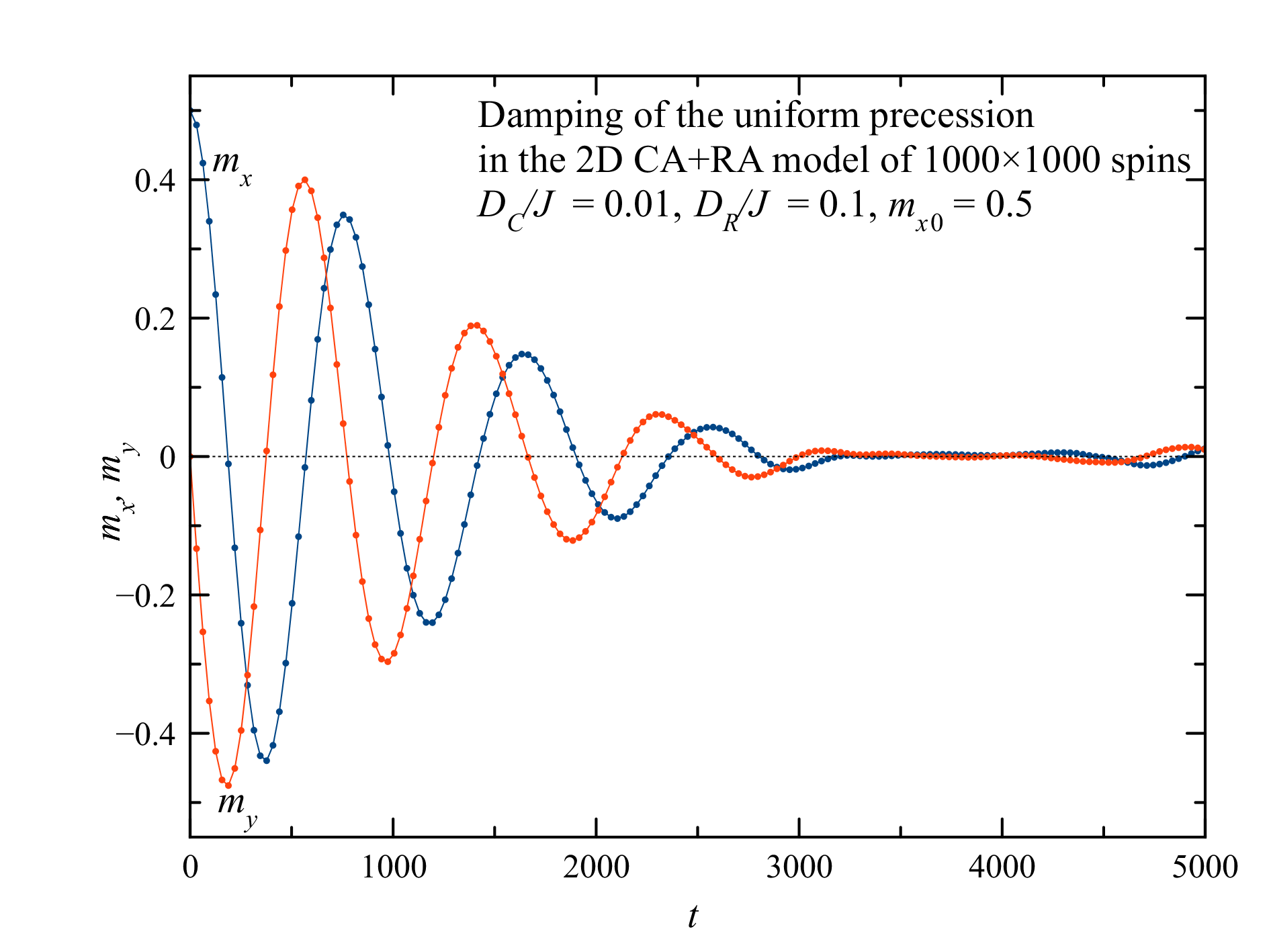}
\par\end{centering}
\centering{}\caption{Damping of the uniform precession of the magnetization. Time dependence
of $m_{x}$ and $m_{y}$ for a large initial deviation, $m_{x0}=0.5$.}
\label{Fig_precession_damping}
\end{figure}
\begin{figure}
\begin{centering}
\includegraphics[width=8cm]{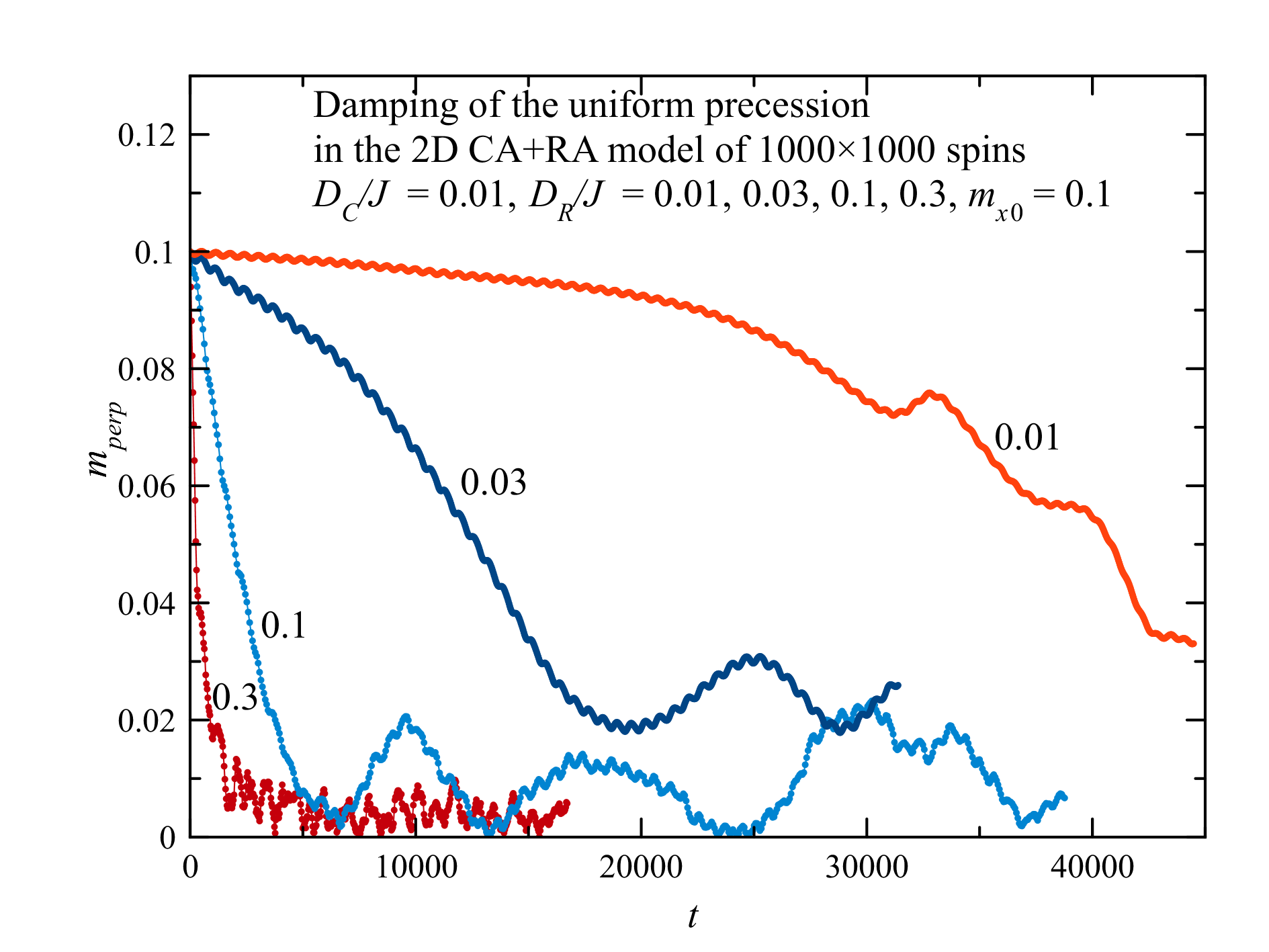}
\par\end{centering}
\begin{centering}
\includegraphics[width=8cm]{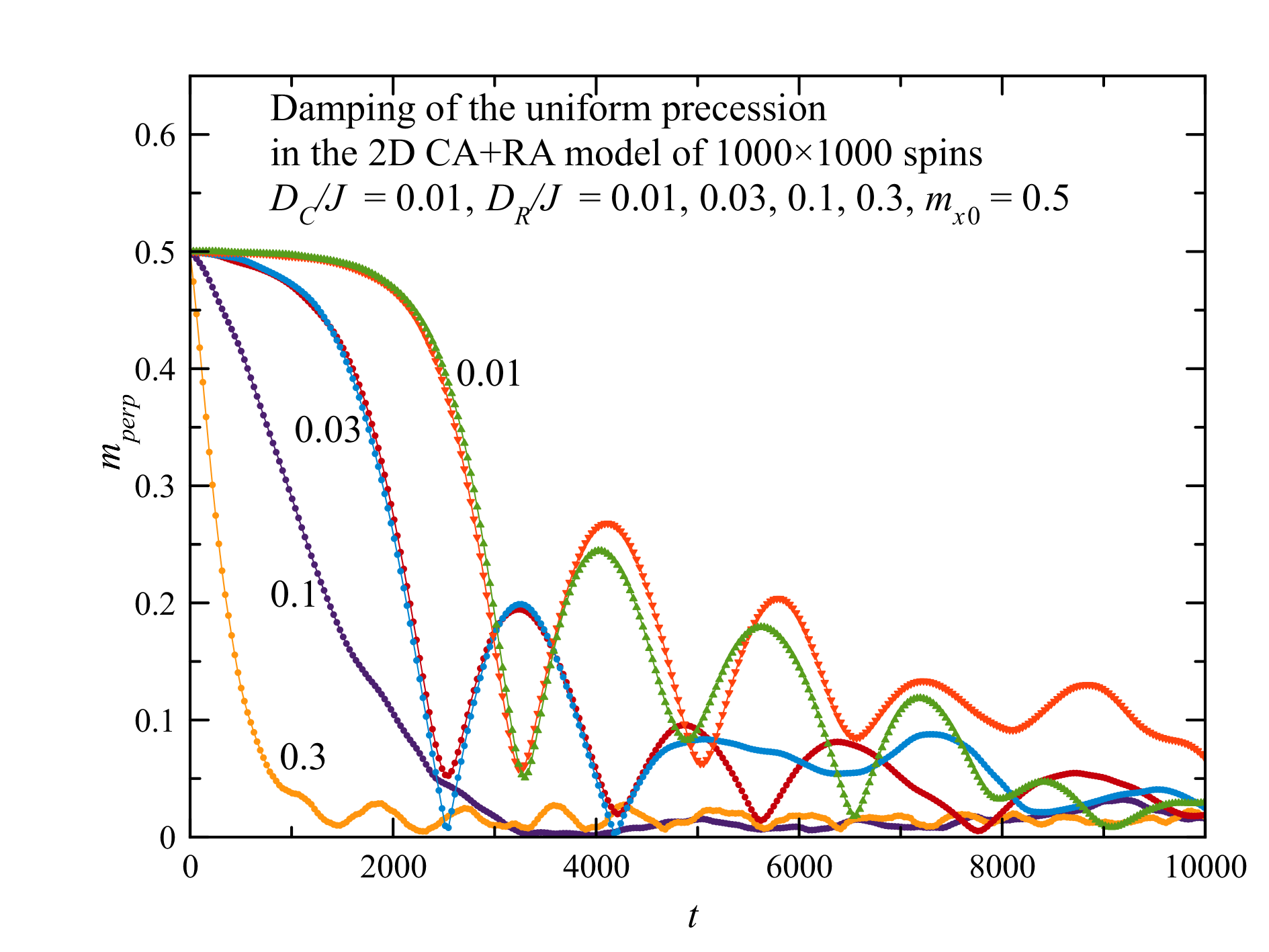}
\par\end{centering}
\caption{Time dependence of the perpendicular component of the magnetization
for three values of the RA. Upper panel: small initial deviation,
$m_{x0}=0.1$. Lower panel: large initial deviation, $m_{x0}=0.5$.}

\label{Fig_mperp_t}
\end{figure}

Close to saturation, the ferromagnetic resonance in the CA+RA model
is located at $\hbar\omega=D_{C}$. Numerical analysis of the oscillations
of the magnetization at that frequency shows a pronounced spatial
localization of spin excitations, see the upper panel in Fig.\ \ref{Fig_localization}.
Here, starting from the equilibrium state $\mathbf{s}_{0}$ at $T=0$
found by the enegy minimization starting from the collinear spin state
along $z$ axis, the system was excited by the uniform magnetic field
of a small amplitude polarized perpendicularly to $z$ axis. The undamped
Landau-Lifshitz equation of motion for classical spins was solved
with the help of the 5th-order Runge-Kutta Butcher's method, and the
square of the spin deviation $\left(\mathbf{s}-\mathbf{s}_{0}\right)^{2}$
was plotted after several periods of the pumpimg. The localization
length was computed as the average width of the local-mode peaks such
as in the upper panel in Fig.\ \ref{Fig_localization}. To compute
this width, we used the $mn$ method (see around Eq. 30 of Ref. \citep{GC-localization}).
Previously, we have demonstrated numerically \citep{GC-PRB2021,GC-localization}
the localization of spin excitations in a pure RA model with $D_{C}=0$,
which must be expected in the random medium. The spatial localization
of the FMR oscillations has been observed in experiments \citep{Suran-1998,McMichael-PRL2003,Loubens-PRL2007,Du-PRB2014}
both away from and close to saturation. In the latter case, however,
when all spins are almost aligned, it is less obvious from the theoretical
perspective.

As is shown in the lower panel of Fig.\ \ref{Fig_localization},
the localization length for spin excitations decreases on increasing
$D_{R}$ as $J/D_{R}$. This agrees perfectly with the analytical
result of the previous section that the localization length is determined
by the ferromagnetic length at $D_{C}=0$ that scales as $J/D_{R}$
in 2D.

\subsection{Decay of the uniform precession due to RA}

\label{Sec_precession_decay}

As excitation modes of the magnet are spatially localized in the presence
of RA, the uniform precession is not a mode of the system and it is
decaying into the local modes. Our numerical analysis shows that in
the presence of RA the uniform precession is damped. Its decay at
$T=0$ after the magnetization $\mathbf{m}=(1/N)\sum_{i}\mathbf{s}_{i}$
of a collinear state was directed at an angle with respect to $z$
axis is shown in Figs.\ \ref{Fig_precession_damping} and \ref{Fig_mperp_t}.
Here, again, we solved the undamped Landau-Lifshitz equation of motion
for classical spins with the help of the 5th-order Runge-Kutta Butcher's
method. At $D_{R}=0$ there is an undamped uniform precession around
$z$ axis. However, as soon as the RA is turned on, uniform precession
decays. As can be seen from in Fig.\ \ref{Fig_mperp_t}, the effect
increases with increasing the RA strength. For small $D_{R}$ and
large initial deviation from $z$ axis, $m_{x0}=0.5$, the decay is
strongly nonlinear and shows partial recurrence.

The oscillating dynamics in the CA+RA model leads to ferromagnetic-resonance
absorption peak near the frequency defined by the core anisotropy,
$\omega\simeq D_{C}/\hbar$, unlike the broadband absorption in the
RA model. The peak is shifted to the left, however, due to the effect
of the temperature and RA, the latter in accordance with Eq. (\ref{frequency}).
The RA and nonzero temperature also contribute to the width of the
microwave absorption peak. Even for $D_{R}=0$, finite temperature
provides the frequency shift and finite line width due to the spin-wave
scattering. 

\section{Conclusions}

\label{Sec_conclusions}

We have studied disordered ferromagnets within random-field and random-anisotropy
models that include the external field and coherent anisotropy. The
latter, even when they are weak compared to their random counterparts,
bring the magnet close to saturation. This is a well-known property
of amorphous ferromagnets which exhibit record low coercive field
and record high magnetic susceptibility. Consequently, for sufficiently
small disordered ferromagnets that do not break into conventional
magnetic domains, the vicinity of saturation is the most interesting
regime of practical importance.

The dependence of the spin-spin correlation function on the parameters
of the model has been computed in this regime for a three-component
spin in two (thin film) and three (bulk material) dimensions. Good
agreement between analytical and numerical results has been obtained.
In the absence of the saturating force (that is, without the external
field $H$ or coherent anisotropy $D_{C}$), the ferromagnetic correlation
length scales inversely as the first power of the disorder strength
in 2D and as the second power in 3D. Close to saturation, the transversal
spin-spin correlations are of interest. We have shown that they decay
exponentially, with the correlation length that scales as $1/\sqrt{H}$
in the random field model with the external field $H$, and as $1/\sqrt{D_{C}}$
in the random anisotropy model with the coherent anisotropy $D_{C}$.

Turning to spin excitations, we have demonstrated that in the vicinity
of the ferromagnetic resonance they are localized regardless of how
close to saturation the system is. Somewhat counterintuitive is the
fact that the localization length is determined by the strength of
the disorder and not by the correlation length of the transverse spin-spin
correlations. This result has been obtained analytically and confirmed
numerically. The disorder is also responsible for the damping of spin
oscillations, which goes up on increasing the disorder strength.

\section*{ACKNOWLEDGEMENT}

This work has been supported by Grant No. FA9550-24-1-0090 funded
by the Air Force Office of Scientific Research.

\section*{Data Availability Statement}

Generated numerical data is available upon reasonable request.

\section*{Author contribution statement}

The authors contributed equally to the work.

\end{document}